# Pseudo-Ring Testing Schemes and Algorithms of RAM Built-In and Embedded Self-Testing

D. Bodean, Gh. Bodean
Radio-Electronics and Telecommunications Faculty
Technical University of Moldova, UTM
Kishinau, R. Moldova
{dianabodean, gbodean}@gmail.com

Wajeb Gharibi
Dept. of Computer Networks,
Jazan University,
Jazan, KSA, Saudi Arabia
gharibi@jazanu.edu.sa

*Abstract*—Scan and ring schemes of the pseudo-ring memory self-testing are investigated. Both schemes are based on emulation of the linear or nonlinear feedback shift register by memory itself. Peculiarities of the pseudo-ring schemes implementation for multi-port and embedded memories, and for register file are described. It is shown that only small additional logic is required and allows microcontrollers at-speed testing. Also, in this article, are given the a posteriori values of some type of memories faults coverage when pseudo-ring testing schemes are applied.

## I. INTRODUCTION

A classical memory built-in self-test (MBIST) scheme contains [1]: (1) a memory BIST *controller*, (2) an *address sequencer* (or *stepper*), (3) a *comparator* for response checking, (4) a *data generator* for inserting test patterns, and (5) a MUX circuit feeding the memory during self-test. The leading position in memory BIST hold the March test algorithms [2]. A March algorithm consists in a set of simple operations such as write, read and compare that are performed iteratively for each memory cell.

Unlike March schemes, the *pseudo-ring testing* (PRT) is based on emulating a linear (or nonlinear) feedback shift register (LFSR) by the memory itself [3]. The idea of pseudo-ring or π-testing is to use a set of memory's cells as the register stages of LFSR and shift this *virtual* register across memory cells. Therefore, it is not the data that are shifting but the virtual LFSR is shifted relatively to data. After shifting via all memory cells, that is called a π-test iteration, a comparison between the (virtual) register final state and the expected one is carried out.

The π-test iteration consists of: *initialization* of virtual automaton, *pushing* this automaton in the space of memory array, *unloading* the automaton final state, and *analysis* of the results The quality of π-testing is estimated by comparing the virtual LFSR final state with expected one. In particular case, when the number of shifting is proportional to period *T* of polynomial $p(x)$, then the comparison with initial state is made: **Init ≥ Fin**.

Π-iteration is a constitutive part of the PRT RAM. The number of π-iterations depends on set of faults to be detected.

Test engineer can define own parameters for each π-iteration. In fact, there are three *controlling parameters* (degree of freedom): LFSR structure, defined by polynomial $p(x)$; initial seed in the π-iteration; addressing mode or trajectory of LFSR.

Pseudo-ring test technique is suitable for a large spectrum of memory devices: one- and multi-port, file registers and cash memory, bit- and word-oriented. Also, a "consistent" running of the virtual LFSR allows at-speed testing of embedded memories, this is important for many complex digital devices such as programmable logic devices (PLD), microprocessors and microcontrollers.

From structural point of view, the π-test scheme follows the classical MBIST architecture, but is not that sophisticated and complex. In this paper the pseudo-ring testing schemes and algorithms will be analyzed. The features of PRT, the file registers and multi-port memories are presented in section 2. Software implementation of the microcontrollers π-testing will be considered in section 3. The paper ends with analysis of PRT fault coverage and concluding remarks.

## II. THE SCHEMES OF PSEUDO-RING TESTING

Compact BIST schemes are based on probabilistic test pattern generation using LFSR [4] and output compaction using signature analyzer [5]. The pseudo-ring testing schemes can be implemented with or without signature analyzer. Π-testing schemes of both cases are presented in Fig. 1. The memory cell in the *ring* scheme plays a role of the virtual feedback stage of LFSR (Fig. 1, a). In the *scan* scheme all register stages connected to feedback are considered as cells of memories (Fig. 1, b).

In the *ring* scheme (see Fig. 1, a) the Signature Analyzer is used to detect some memory faults that can be "omitted" by virtual LFSR during a π-iteration. It is easy to see, the Signature Analyzer can be an external one. In the *scan* test mode the shift register ShReg (see Fig. 1,b) is only used for temporary data storage. So, ShReg-unit is the copy of the virtual register of LFSR, emulated by cells of memory. When the number of RAM ports is equal to register length, there is no need to use the register ShReg.

The signal Read/Write is generated, e.g. by GenA, in correspondence with RAM specification. At each clock of time the virtual register is shifted relatively to data. Address generator or sequencer GenA determines the trajectory of LFSR shifting. Three types of trajectories can be selected: *counting up*, *counting down* or *pseudorandom*. It's necessary to outline that in some works (for example, [6, 7]) is mentioned, that use of different initial conditions such as address order or background changing can increase the test quality of the March algorithm.

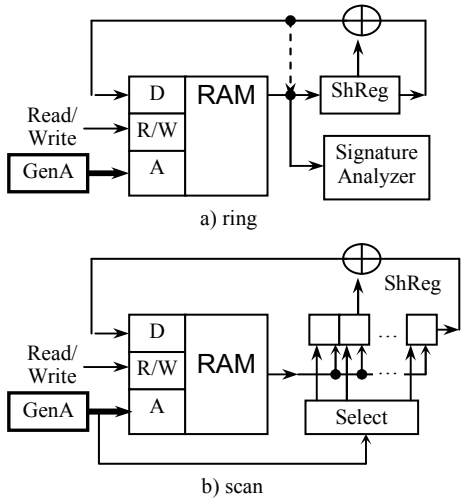

Figure 1. Schemes of Π-testing.

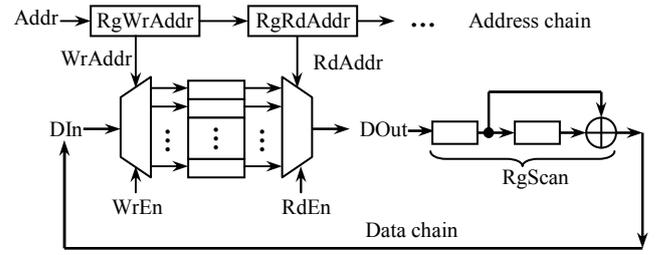

Figure 2. Π-testing scheme of the registers file.

In addition to the three parameters, specified in the Introduction, to control the π-testing can be used another, fourth, parameter namely the *input* or *input-output inversion* of data. Inversion of data allows extending the variety of automaton states, which are not specific for classical LFSR. For example, double inversion (input-output) of data allows crossing LFSR through states 11…1 (full in the memory array by ones). As a consequence the π-iteration can simulate the March algorithms.

Tested memory can be either bit- or word-oriented. PRT scheme, i.e. automaton, is a classical LFSR that process data over Galois field **GF**(2), when RAM is bit-oriented. Automaton is a so-called General LFSR [8] that processes data over extended **GF**$^k$($2^m$), if RAM is word-oriented, where $m$=deg $p(x)$ is the cell size and $k$=deg $q(z)$ is the degree of irreducible polynomial over **GF**$^k$($2^m$). Obviously, the value $k$ is multiple to the size $N$ of memory array. In following subsections the specifics of PRT schemes implementation for different memory type are presented and described.

### A. Pseudo-Ring Testing of Small Embedded RAM

In Fig. 2 is shown the scan π-testing scheme of a register file. Symbol ⊕ specifies a logical scheme of XOR-gates, that depends on selected LFSR type: (1) over extended Galois field **GF**$^n$($2^m$), where $m$ is the size of cell, $n$ is the number of register stages; (2) group of $n$ (*homo*- or *hetero*geneous) LFSR over **GF**($2^m$), where $n·m$= cell size. Selection of LFSR type (1) or (2) depends on hypothesis about faults: *intra*- or *inter*-words.

For the first LFSR type the hardware overhead will be equal to $m$ 2-inputs XOR-gates. For the second LFSR type hardware overhead will be the same as for LFSR (1) plus some XOR-gates that implement multiplication by a constant over corresponding field **GF**$^n$($2^m$). In both cases registers RgWrAddr and RgRdAddr are indispensible parts of the scan chain. Remark that RgScan can be built-in, as well as external to unit under test.

Π-testing is defined as follows: by computer-aided design tools [9] the test sequences are generated, simulated and verified for a prescribed list of faults. Further, the prepared tests are feed to Address and Data chain inputs. The Address and Data chains are synchronized separately (synchronizations inputs are not shown in Fig. 2). This feature allows to schedule π-testing to detect various faults types. For example, as was shown in [9], the corresponding control and configuration of π-testing scheme allows detecting all static single- and two-cell faults and all dynamic single-cell faults in the period of time proportional to 54$N$, where $N$ is the RAM array size.

### B. Pseudo-Ring Testing of Multi-Port Memory

Most of the multi-port memory circuits are word-oriented. Two-port memory will be used further to illustrate the synthesis of PRT scheme. Among all possible designs of PRT for two-port RAM, in [10] are selected two most attractive from hardware implementation standpoint.

The initial data of designing a pseudo-ring system are: size $m$ of memory cell, size $N$ of array memory array, polynomial $q(x)$ over extended Galois field **GF**$^k$($2^m$), where $k$= deg $q(x)$ is the number of register stages. Period $T$ characterizes the GLFSR behavior. If $q(z)$ is a primitive irreducible polynomial then period is maximal: $T$= $(2^m)^k$–1. In the case $k$= 2 a transition of the virtual GLFSR over **GF**($2^4$) is shown in Fig. 3.

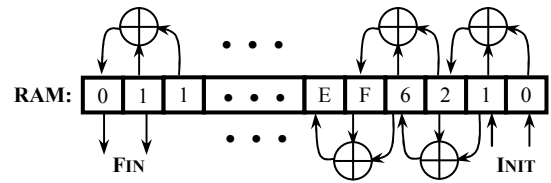

Figure 3. Diagram of the pseudo-ring testing with GLFSR.

Test iteration, shown in Fig. 3, has an equivalent description in memory test language: $\Updownarrow \{r_i, r_{i+1}, w_{i+2}(r_i \oplus r_{i+1})\}$, where {…} signifies the iterative performing of the included operation, $w_{(·)}$ is the write to address (·), $r_{(·)}$ is the read from address (·), and symbol ⊕ signs the sum *modulo q(z)*. For example, the polynomial $q(z)$= 1+2$z$+2$z^2$ is one of the primitive irreducible polynomials over **GF**$^2$($2^4$) with *field generator polynomial $p(x)$*= 1+$x$ +$x^4$ over **GF**(2). So, for this example is needed to multiply modulo $q(z)$ by 2 the two adjacent cell's values $z$ and $z^2$ and sum *modulo* $p(x)$, i.e. XOR the resulted values. It is well known that multiplication with a constant over Galois field is implemented by a combinational circuit [11]. Thus, the operation sum *modulo* $q(z)$ is accomplished by not "costly" logical circuit.

To facilitate π-testing a modification of the standard two-port (A and B) memory architecture is proposed. This modification needs: (1) allowing the "conversion" of the existent address registers RgAddr to counters and (2) additional hardware overhead of specific XOR-logic on RAM chip area, i.e. block ⊕. Generic VHDL entities were elaborated to generate the corresponding units of π-testing scheme for bit-oriented as well as for word-oriented RAMs. The supplementary hardware overhead is negligible. Table 1 shows the rate of hardware overhead when designing a π-test system.

TABLE I. Π-TESTING SCHEME HARDWARE OVERHEAD

| Memory Array Size | LFSR | | | | | GLFSR | | | | |
|---|---|---|---|---|---|---|---|---|---|---|
| | *1 kb* | *32 kb* | *1 Mb* | *32 Mb* | *1 Gb* | *1 kB* | *32 kB* | *1 MB* | *32 MB* | *1 GB* |
| Hardware overhead, % | 3,4 $10^{-4}$ | 1,5 $10^{-5}$ | 6,2 $10^{-7}$ | 2,4 $10^{-8}$ | 8,0 $10^{-10}$ | 7,3 $10^{-5}$ | 2,86 $10^{-6}$ | 1,07 $10^{-7}$ | 3,9 $10^{-9}$ | 1,4 $10^{-10}$ |

*b* means bits and *B* means bytes.

Considering estimation $O(\pi\text{-iteration}) = N$, the π-test length is of order $O(r \cdot N)$, where $r$ is the number of π-iterations. In [3] was proved that for hard-to-detect, namely single, faults the optimal number of π-iterations is bounded below by value $k+1$. It up to the test-engineer only to find the optimal parameters of PRT for a prescribed set of RAM faults. Analyzed in this section π-testing schemes allow extending the pseudo-ring test technique for embedded memory of microcontrollers.

### III. PSEUDO-RING TESTING FOR MICROCONTROLLER

The application note [12] describes the Cyclic Redundancy Check (CRC) based algorithm for testing the program memory of AVR RISC microcontrollers. The article [13] focuses on adaptation of March bit-oriented algorithm for at-speed BIST the Atmel AVR-controllers. In this section the peculiarities of the implementation of pseudo-ring AVR-microcontrollers self-testing is described.

LFSR's structure will be "assembled" by the microcontroller registers, the polynomial algebra operations – multiplication and addition, will be performed by using built-in flash EEPROM memory. To read the contents of flash memory cells will use the LPM instruction (loap program memory) that load a data byte from the FLASH program memory into the register file. The Z-register in register file is used to access the program memory.

All operations are executed in the extension of Galois field $\mathbf{GF}^k(2^m)$ with coefficients of residue classes of polynomials modulo $p(x)$. Next will be described an example of implementation the virtual GLFSR defined by irreducible polynomial $q(z)=1+z+9z^2$ over $\mathbf{GF}(2^4)$ with generator polynomial $p(x)=1+x+x^4$ over $\mathbf{GF}(2)$. The sum $z+9z^2(\mathbf{mod}\, p(x))$ table is needed to generate before implementing modular operations. The resulted decimal values are saved in the first 256 bytes of data memory (see Fig. 4).

```
.eseg     ; org    0
.db    0,1,2,3,4,5,6,7,8,9,10,11,12,13,14,15
.db    9,8,11,10,13,12,15,14,1,0,3,2,5,4,7,6
.db    1,0,3,2,5,4,7,6,9,8,11,10,13,12,15,14
.db    8,9,10,11,12,13,14,15,0,1,2,3,4,5,6,7
.db    2,3,0,1,6,7,4,5,10,11,8,9,14,15,12,13
.db    11,10,9,8,15,14,13,12,3,2,1,0,7,6,5,4
.db    3,2,1,0,7,6,5,4,11,10,9,8,15,14,13,12
.db    10,11,8,9,14,15,12,13,2,3,0,1,6,7,4,5
.db    4,5,6,7,0,1,2,3,12,13,14,15,8,9,10,11
.db    13,12,15,14,9,8,11,10,5,4,7,6,1,0,3,2
.db    5,4,7,6,1,0,3,2,13,12,15,14,9,8,11,10
.db    12,13,14,15,8,9,10,11,4,5,6,7,0,1,2,3
.db    6,7,4,5,2,3,0,1,14,15,12,13,10,11,8,9
.db    15,14,13,12,11,10,9,8,7,6,5,4,3,2,1,0
.db    7,6,5,4,3,2,1,0,15,14,13,12,11,10,9,8
.db    14,15,12,13,10,11,8,9,6,7,4,5,2,3,0,1
```

Figure 4. Table of $9z^2+z$ **mod** $(1+x+x^4)$.

Algorithm of checksum calculation will run as follows: Beginning with the first cell of program memory with address $i$=0, clock by clock the content of the two GLFSR stages are summed modulo $q(z)$, and the result is XOR-ed with value of the $i$-th memory cell. The GLFSR is shifted, so the less significant word (LSW) is moved in the most significant word (MSW), and the result, obtained in previous calculus, is saved in the LSW stage. The corresponding listing of AVR-subroutine of the above algorithm is shown in Fig. 5.

```
; ***** Subroutine Register Variables
.def    sizeL    = r17           ; Program code
.def    sizeH    = r18           ; size register
.def    LSW      = r19           ; Lower byte of GLFSR
.def    MSW      = r20           ; Upper byte of GLFSR

PiSign: ldi     sizeL, low(end_P+1) ; Load end of
        ldi     sizeH, high(end_P+1); program memory address
        clr     zL              ; Clear Z pointer
        clr     zH
_pi:    cp      zL, sizeL       ; Check for end of code
        cpc     zH,sizeH
        brge    piEnd           ; Jump if end of code
        out     EEARL, MSW      ; Output address low
        out     EEARH, Zero     ; and high byte
        sbi     EECR, EERE      ; Set EEPROM read strobe
        mov     MSW, LSW        ; MSW← LSW -- shift GLFSR
        in      r0, EEDR        ; r0 ← az^2
        eor     LSW, r0         ; LSW ← LSW ⊕ r0 {az^2+z}
        lpm                     ; r0 ← Code[ i]
        eor     LSW, r0         ; LSW ← LSW ⊕ r0
        adiw    zL, 1           ; next cell i of
        rjmp    _pi             ; Code Memory
piEnd:  ret                     ; from PiSign8
```

Figure 5. Subroutine of π-testing the AVR-controller program memory.

The subroutine **PiSign** is called from main program as listed below.

```
.include "8515def.inc"
;***** Constants
.equ    end_P_ = 0x1FFF         ;Size of program memory (bytes)
;***** Register Variables
.def    Zero     =r14           ; constant zero register
.def    sum      =r15           ; CRC checksum
.def    temp     =r16           ; temporary register
.def    glfsr    =r17           ; linear feedback shift register

;//////////////// Program start – execution starts here //////////////////////
.cseg
.org    $0000
        rjmp RESET              ;Reset handle
.org    11
```

```
;//////////// Starts of Main Program //////////////////////
Reset:  ldi   temp, high(RAMEND)  ; Initialize stack pointer
        out   SPH,temp            ; High byte only required if
        ldi   temp,low(RAMEND)    ; RAM is bigger than 256 Bytes
        out   SPL, temp
        clr   Zero                ; set constant 0
        clr   glfsr               ; reset GLSFR register
        rcall PiSign              ; get GLFSR value
        ; Output GLFSR value to PortA
.exit
```

About 385000 cycles are required to run the program outlined above. The elaborated program is a draft aimed to run on the simulator. Therefore, it does not take into account peculiarities of read and write in the EEPROM of various members of the AVR family. Based on described algorithm of π-testing EEPROM-memory of the AVR-controller, we have also developed algorithms and corresponding programs to PRT BIST other types of memory of the microcontroller.

Speed related faults detection is one of the aims of embedded built-in self-test program. A way to detect these faults is by using back-to-back (BtB) memory cycles. To provide this "desideratum" one must follow the BtB recommendations contained in the [13], but adapted to the psedo-ring testing. In addition to at-speed testing of microcontroller, also arouse interest the testing of static faults.

## IV. Fault Coverage of the Pseudo-Ring Testing

In this section the results of simulation the trivial π-testing are presented. The number of π-iterations in the test experiment is equal to $m+1=3$ and start with significant seeds. This π-test experiment was performed for the list of 31 single and 86 two-cell faults, proposed for one-port SRAM in [14]. For PRT simulation were applied the tools described in [9], and the ring scheme shown in Fig. 1, a) was used as a π-testing scheme. The results of this experiment have showed that the average fault coverage $R$ of single cell faults is equal to 0.9128 and for the two-cell faults is equal to 0.8627. Also, remark that in the class of single cell faults the most difficult to detect is the Write Destructive fault, and in the class of two-cell faults - the Transition Coupling fault (with $R = 0.75$ both).

Underline that the obtained results are "reliable" for one-bit oriented memories with arbitrary array size. As can be seen from Table II, there are such type of faults for which the estimation $R$ of LFSR is higher than estimation $R$ of SA, and vice versa. It is necessary to mention that for memory chips more than 150 of possible faults are known [14].

## V. Conclusions

Pseudo-ring testing (PRT) is a new technique to built-in self-testing of different type of memory circuits, and to (embedded) self-testing the memories of microcontroller units (MCU) and (micro)processors. The PRT or π-testing is based on emulation of a linear automaton such as linear feedback shift register by memory itself. Therefore, rich theory of linear automaton can be utilized to solve pressing BIST problems. As a result, test-engineers get a powerful methodological tool to organize, control and manipulate the RAM test procedure.

Two basic schemes – ring and scan, of π-testing are presented in this paper. Relative to the memory chip the PRT-schemes can be implemented externaly, internaly or mixed. In all cases, a few hardware overhead are needed for scan or ring scheme implementation. In some cases, just an extension of inbuilt memory components abilities may be sufficient.

The proposed schemes are suitable both for bit-oriented as well as for word-oriented memories and provide adequate architecture support to allow interfacing with known BIST standard, e.g. IEEE 1149. Another remarkable property of the π-testing, that must be noted, is the *invariability* of the testing scheme. It means that the same π-test scheme can be applied (without essential adjustments) as for single-port so for multi-port memories. Four control parameters (as degree of freedom) are "subject" to the test-engineer for synthesis a fast and high fault coverage π-test.

One of the distinct features of the described PRT in this paper is that the quality estimation of π-testing is performed at the end of PRT by comparing the final of the emulated automaton with the expected one. This feature allows at-speed testing, which is also important for microcontrollers' embedded testing. An example of AVR-controller embedded π-testing is shown in this article. The example is implemented in assembler language and is about 40% shorter than the known Cyclic Redundancy Checking ATMEL-program.